\documentclass[11pt]{article}
\usepackage{latexsym}

\usepackage{amssymb}
\usepackage{amsmath}

 \usepackage{epsfig}
 \usepackage{graphicx}

\newcommand{\be}{\begin{equation}}
\newcommand{\ee}{\end{equation}}
\newcommand{\ba}{\begin{eqnarray}}
\newcommand{\ea}{\end{eqnarray}}
\newcommand{\baa}{\begin{array}}
\newcommand{\eaa}{\end{array}}
\newcommand{\bi}{\begin{itemize}}
\newcommand{\ei}{\end{itemize}}
\newcommand{\edoc}{\end{document}}

\newcommand{\nn}{\nonumber \\}
\newcommand{\nr}[1]{(\ref{#1})}
\newcommand{\la}[1]{\label{#1}}

\newcommand{\rmi}[1]{{\mbox{\scriptsize #1}}}
\newcommand{\fr}[2]{{\frac{#1}{#2}\,}}
\newcommand{\fra}[2]{{\textstyle{\frac{#1}{#2}\,}}}  

\newcommand{\bfx}{{\bf x}}

\newcommand{\bfk}{{\bf k}}

\def\tr{{\rm Tr\,}}

\def\im{{\rm Im\,}}

\def\CL{{\cal L}}

\def\CN{{\cal N}}
\def\CO{{\cal O}}
\def\gsim{\raise0.3ex\hbox{$>$\kern-0.75em\raise-1.1ex\hbox{$\sim$}}}
\def\lsim{\raise0.3ex\hbox{$<$\kern-0.75em\raise-1.1ex\hbox{$\sim$}}}

\begin{document}

\begin{titlepage}
\begin{flushright}
HIP-2011-09/TH\\
 BI-TP 2011/11\\
\end{flushright}
\begin{centering}

\vfill
{\Large{\bf Frequency and wave number dependence of the shear
correlator in strongly coupled hot Yang-Mills theory}}

\vspace{0.8cm}

\renewcommand{\thefootnote}{\fnsymbol{footnote}}

K. Kajantie$^{\rm a,b}$\footnote{keijo.kajantie@helsinki.fi},
Martin Kr\v s\v s\'ak$^{\rm c}$\footnote{krssak@physik.uni-bielefeld.de},
M. Veps\"al\"ainen$^{\rm a}$\footnote{mikko.vepsalainen@helsinki.fi},
Aleksi Vuorinen$^{\rm c}$\footnote{vuorinen@physik.uni-bielefeld.de}
\setcounter{footnote}{0}

\vspace{0.8cm}

{\em $^{\rm a}$%
Department of Physics, P.O.Box 64, FI-00014 University of Helsinki,
Finland\\}
{\em $^{\rm b}$%
Helsinki Institute of Physics, P.O.Box 64, FI-00014 University of
Helsinki, Finland\\}
{\em $^{\rm c}$%
Faculty of Physics, University of Bielefeld, D-33501 Bielefeld, Germany\\}
\vspace*{0.8cm}

\end{centering}

\noindent
We use AdS/QCD duality to compute
the finite temperature Green's function $G(\omega,k;T)$ of the shear operator $T_{12}$
for all $\omega,k$ in hot Yang-Mills theory. The goal is to assess how the existence of scales like
the transition temperature and glueball masses affects the correlator computed in
the scalefree conformal $\CN=4$ supersymmetric Yang-Mills theory. We observe sizeable
effects for $T$ close to $T_c$ which rapidly disappear with increasing $T$. Quantitative
agreement of these predictions with future lattice Monte Carlo
data would suggest that QCD matter in this
temperature range is strongly interacting.

\vfill \noindent


%

\vspace*{1cm}

\noindent


\vfill

\end{titlepage}

\section{Introduction}
There is a standard framework for computing Green's functions
in supersymmetric $\CN=4$ Yang-Mills theory using AdS/CFT duality
\cite{policastrostarinets,ss,kovtunstarinets,
kovtunstarinets2,teaney,gubserpufu,romatschkeson}.  In \cite{kv}, this was carried out
explicitly over the entire $\omega,k$ plane for an operator coupling
to a scalar field. The specific goal was to compute the real static ($\omega=0$)
Green's function both directly and by integrating over $\omega$. The result
was then Fourier transformed to a spatial Green's function and compared with
a computation
in next-to-leading order QCD perturbation theory \cite{lvv1,lvv2}, see also
\cite{meyer,iqbalmeyer}.

The previous calculation was carried out in a scalefree theory. The purpose
of this article is to see what effects arise from $\Lambda_\rmi{QCD}$, i.e.,
scale invariance breaking in SU($N_c$) Yang-Mills theory. Physically this
manifests itself in the existence of glueballs and a first order
phase transition\footnote{A computation in a non-conformal theory but without
a phase transition has been carried out in \cite{springer}.}.
The predictions are quantitative and it is suggested that
their comparison with lattice Monte Carlo data may give a better handle on
verifying the strongly interacting nature of QCD matter in the temperature range
$T_c...10T_c$. There is definite phenomenological evidence for this (magnitude
of pressure, small shear viscosity, jet quenching), but no definite nonperturbative
test \cite{iancumueller}. In fact,
how can the system be strongly interacting
while so many lattice results \cite{karschetal} agree with ideal gas?
One concrete test \cite{teaney,schaferteaney} would be to verify
if the shear spectral function has a transport peak at $\omega=0$ or not;
our calculation addresses this question.

There are many candidate gravity duals of SU($N_c$) Yang-Mills theory. The dual
we want to apply is improved holographic QCD (IHQCD)
as developed in \cite{kiri1,kiri2,kiri3,kiri4}.
This is a bottom-up Anti de Sitter (AdS) gravity+dilaton model which 
combines elegantly asymptotic
freedom in the ultraviolet (UV) with a given beta
function and confinement in the infrared (IR).
IHQCD has to be solved numerically and we can so far only compute the imaginary
part of the Green's function the case $k=0$ as a function of $\omega$. 
Since we want to treat the entire $\omega,k$ plane,
and for increased transparency, 
we shall in this note also use a simplified version of IHQCD for which
the gravity background is analytically known \cite{kiri2}. It
contains qualitatively the relevant features of IHQCD and
computations in the overlapping region confirm that it gives
good idea of what one can expect of scale invariance breaking effects on
Green's functions in SU($N_c$) Yang-Mills theory.

IHQCD contains the background metric and the scalar dynamically coupled, and
fluctuation analysis should be carried out taking this into
account \cite{kiri_gravfluct,springer1,springer2}.
The transverse tensor mode ($T_{12}$ with $\bfk=(0,0,k)$; also called
shear mode or scalar mode) does not couple to other
ones and we thus consider only this case and study the retarded Green's function
\be
G(\omega,k;T)=\int_{-\infty}^\infty dt\,d^3x\,e^{i\omega t-ikx^3}
\langle\, i\theta(t)[T_{12}(t,\bfx),T_{12}(0,{\bf 0})]\,\rangle_T
=\int_{-\infty}^\infty{d\omega'\over\pi}{\rho(\omega',k)\over\omega'-\omega-i\epsilon}.
\la{G}
\ee

On scales and notation: in IHQCD, the QCD-like scale $\Lambda$ arises as an integration
constant of the dynamical dilaton field equation, giving the $z$ dependence of the dilaton
field, $z$ being the fifth dimension. In the model studied here, it is introduced as a constant
of dimension 1 by the $z$ dependence of the ansatz \nr{bmodel}.  The
spin 2 glueball spectrum (squared) then will be $4\Lambda^2(2+n)$, $n=0,1,2,..$ (analytical),
and the transition temperature $T_c=0.400\Lambda$ (numerical result).
Energy (and wave number) can be made dimensionless
by scaling with either $T$ or $\Lambda$ and we shall define
\be
\omega=\pi T\,\tilde\omega=2\Lambda\,\hat\omega\approx5.00\,T_c\,\hat\omega,
\la{enscales}
\ee
similarly for $k$. As the dimensionless fifth coordinate we shall use
\be
y=\Lambda^2 z^2.
\la{defy}
\ee
Unless specifically noted, results for the retarded Green's function $G$ are given
without writing explicitly the dimensionless factor $\CL^3/(4\pi G_5)$, where $\CL$ is the
AdS radius and $G_5$ the 5dim gravitational constant.

\section{Gravity dual}
\subsection{IHQCD}
IHQCD is based on the gravity + scalar action, in standard notation,
\be
S={1\over16\pi G_5}\left\{\int d^5x\,\sqrt{-g}\left[R-\fra43(\partial_\mu\phi)^2+V(\phi)\right]-
\int d^4x\,\sqrt{-\gamma}\,\biggl[2K+{6\over\CL}+{\CL\over2}R(\gamma)\biggr]\right\}.
\la{S}
\ee
The background, with boundary at $z=0$, is chosen to be
\be
ds^2=b^2(z)\left[-f(z)dt^2+d\bfx^2+{dz^2\over f(z)}\right],
\la{bg}
\ee
where $b(z),f(z),\phi(z)$ are solutions of the first order system
\ba
\dot W&=& 4bW^2-\fra{1}{f}(W\dot f+\fra13 b V),\quad W= -\dot b/b^2,\la{eq1}\\
\dot b&=& -b^2W,\nn
\dot \lambda&=&\fra32\lambda\sqrt{b\dot W},\la{eqlambda}\\
\ddot f&=&3\dot fbW,
\la{eqf}
\ea
(see Appendix A of \cite{kiri3}). The coupling of the boundary theory is associated with
$\lambda(z)=e^{\phi(z)}$, and a specific combination of the fields, arising in connection
of solving Einstein's equations ($\dot b=db/dz$),
\be
\beta(\lambda)={\dot\lambda\over\dot b/b},
\la{bfn}
\ee
is associated with the beta function of the boundary theory. An important general
property of the solutions is that $b$ and $f$ are monotonically decreasing and $\lambda$
is monotonically increasing as functions of $z$.

The solutions have to satisfy the following properties. In the UV, $z\to0$,
\be
b(z)\to{\CL\over z}.
\la{bL}
\ee
Imposing asymptotic freedom to leading order, one then has,
\be
\beta(\lambda)={\dot\lambda\over\dot b/b}=-b_0\lambda^2=-z{d\lambda\over dz},
\ee
implying that the leading term of $\lambda(z)$ near the boundary is
\be
b_0\lambda(z)={1\over L},\quad L\equiv \log{1\over\Lambda z},
\la{laL}
\ee
$\Lambda$ = constant of integration. This is how the dimensional transmutation of
SU($N_c$) YM theory appears in the gravity dual. Note that on the boundary
$\lambda\to0$ but $\phi\to-\infty$.

Including more powers of $\lambda$ in the beta function, one can derive expansions
in $\lambda$ for all the fields. For the record, we have collected some of them in
Appendix A.

In the IR, for $z\to\infty$, the crucial confinement criterion is that out of
the many different solutions of Einstein's equations \nr{eqf} one should choose
the one satisfying
\be
\beta(\lambda)=-\fr32\lambda\biggl[1+{3(\alpha-1)\over4\alpha}{1\over\log\lambda}+
\CO\biggl({1\over\log^2\lambda}\biggr)\biggr].
\ee
The parameter $\alpha>1$ here is connected with the IR behavior of all the other
background functions so that, for example,
\be
b(z)\to e^{-(\Xi z)^\alpha} (\Xi z)^p,\quad \Xi={\rm number}\times\Lambda,\quad p={\rm number}.
\la{bas}
\ee
The appropriate IR relations are summarised in Appendix B. We shall next choose $p=-1$ to
extend the form to all $z$ and $\alpha=2$ to have a glueball spectrum with 
$m^2\sim{\rm integer}$ \cite{son_linconf}.

\subsection{Model: approximate version of IHQCD}\la{subsectmodel}
Our model for simulating the essential properties of IHQCD is defined by
a simple ansatz \cite{kiri2} for $b(z)$:
\be
b(z)={\CL\over z}\exp(-{\fra13}\Lambda^2z^2)={\CL\Lambda\over \sqrt{y}}\exp(-\fra13 y),
\la{bmodel}
\ee
where we often set
$\CL=1$ and where we use the dimensionless distance variable $y$ in \nr{defy}.

Starting from \nr{bmodel} one can derive all the other bulk fields.
First, integrating \nr{eqlambda} leads to
\be
\phi'(y)={\lambda'(y)\over\lambda(y)}=\fr12\sqrt{1+{9\over2y}},
\ee
which integrates to
\be
\lambda(y)/\lambda_0=\exp(\fr12 \sqrt{y(y+\fra92})\,
\left(\sqrt{y}+\sqrt{y+\fra92}\right)^{9/4}.
\la{lay}
\ee
Now that $\lambda(y)$ is known, the $\beta$ function as specified by IHQCD is obtained:
\be
\beta(\lambda)={\lambda\phi'(y)\over b'(y)/b(y)}=-\fr32\lambda{\sqrt{1+9/(2y)}\over 1+3/(2y)}.
\la{betaf}
\ee
In the IR this goes like
\be
\beta(\lambda)=-\fr32\lambda\biggl(1+{3\over4y}+...\biggr)=
-\fr32\lambda\biggl(1+{3\over8\log\lambda}+...\biggr)
\ee
and thus satisfies explicitly the IHQCD confinement criterion.
In the UV, $\lambda(y)$ goes to $\lambda(0)(1+3\sqrt{y/2}+9y/4+...)$, i.e.,
this ansatz does
not produce a $\lambda(y)$ vanishing logarithmically at the boundary. The far UV region is
inessential for the treatment of thermodynamics in IHQCD \cite{aks} and the absence of the
"spurious logarithms" \cite{kirispurious} simplifies the AdS analysis significantly.

To get $f(z)$, we note that \nr{eqf} integrates to ($C$ is a constant)
\be
\dot f(z)=C/b^3(z).
\la{fdoteq}
\ee
The greatest utility of the ansatz \nr{bmodel} lies in that this is integrable
in closed form. Using $f(0)=1,\,\,f(y_h)=0$, one has
\ba
f(y)&=&1-{(y-1)e^y+1\over (y_h-1)e^{y_h}+1}={y_h-1-(y-1)e^{y-y_h}\over y_h-1+e^{-y_h}}\nn
&=&{1\over y_h-1+e^{-y_h}}\sum_{n=1}^\infty
{y_h+n-1\over n!}(-1)^{n+1}(y_h-y)^n.
\la{fy}
\ea
Finally,
the scalar potential $V(\lambda)$ can be solved from \nr{eq1} with the answer
(keeping $\CL$ here to show where the dimensions of $V$ come from)
\be
V(\lambda(y))={12\over\CL^2}\cdot e^{\fra23 y}
\bigl[(\fra13 y^2+\fra56 y+1)f(y)-(\fra12+\fra13 y)yf'(y)\bigr].
\la{Vcalc}
\ee
In the vacuum $f=1,\,f'=0$ and, using \nr{lay}, the potential at large $\lambda$ is seen to
behave as
\be
V(\lambda(y))={12\over\CL^2}\lambda^{4/3}\sqrt{\log\lambda}(1+\CO(1/\log\lambda)),
\ee
again in agreement with the confining potential criteria in IHQCD. A conceptual problem
with $V(\lambda)$ in \nr{Vcalc} is its dependence on $y_h$. However, what matters for
solving the background is the range $0<y<y_h$, and in this range
the potentials computed from \nr{Vcalc} for $f=1$ and $f=f(y)$ are very close to each other.
They differ maximally at $y=y_h$ and this difference is maximally 4.7\% if $y_h=0.79$.

\subsubsection{Thermodynamics}
To assess the relevance of the model $b(z)$ in \nr{bmodel}, we first work out its thermodynamics.
From $4\pi T=-\dot f(z_h)$ and \nr{fy}, one obtains the entire thermodynamics parametrised
by $y_h$ (including now the factor $\CL$ in $b(z)$):
\ba
T(y_h)&=&{\Lambda\over2\pi}\,{y_h^{3/2}\over y_h-1+e^{-y_h}}\equiv 2\Lambda\,\hat T,\la{tyh}\\
s(y_h)&=&{\CL^3\over4G_5}\Lambda^3 \,y_h^{-3/2}e^{-y_h},\\
p(y_h)&=&\int_{y_h}^\infty dx\,[-T'(x)]s(x).
\ea
Using \nr{lay}, we may replace $y_h$ by $\lambda_h$, which is used in \cite{kiri2}.
The overall structure of $T$ is such that it diverges
both for small $y_h$ ($\pi T\to\Lambda/\sqrt{y_h}$, the big black hole limit)
and for large $y_h$ ($\pi T\to\fra12\Lambda\sqrt{y_h}$, the small BH limit). Thus,
$T(y_h)$ has in between a minimum which lies at
\be
y_h=2.149,\qquad T_\rmi{min}=0.3962\Lambda.
\ee
For $T>T_\rmi{min}$ the system is in a deconfined plasma phase with pressure $\sim N_c^2$,
but actually in a part of this range, $T_c>T>T_\rmi{min}$, the plasma phase is metastable.
The stable phase for 
$0<T<T_c$ is a thermal glueball gas phase with pressure $\sim N_c^0$, i.e.,
in this approach $p=0$. The transition temperature $T_c$ is where the pressure
vanishes, which happens at
\be
y_h=y_c=1.6863,\qquad T_c=0.4000\Lambda=1.0096T_\rmi{min}.
\la{tc}
\ee
In the UV, where $y_h\to0$ and $\pi T\to\Lambda/\sqrt{y_h}$,
\be
{p\over T^4}\to {\CL^3\over4G_5}{\pi^3\over 4},
\la{UVval}
\ee
independent of the IR scale $\Lambda$. This can be used to fix the dimensionless
combination $\CL^3/(4 G_5)$, which then becomes $\sim N_c^2$. For example,
in the present AdS/QCD context one may match at some large  
$T\gg T_c$ to the pressure of weakly interacting
gluon gas, $p/T^4=\pi^2N_c^2/45$. Weakly interacting $\CN=4$ supersymmetric YM matter
has $p/T^4=\pi^2N_c^2/6$ while the strongly interacting one has $p/T^4=\pi^2N_c^2/8$ with
\be
{\CL^3\over4\pi G_5}={N_c^2\over2\pi^2}.
\la{Lthree}
\ee

The resulting thermodynamics is plotted in Fig.~\ref{modelth}. The
interaction measure is broader than that of pure YM at large $N_c$ and the
latent heat smaller. The latent heat is related to the large $T$ asymptotics by
\be
L={\epsilon_c\over T_c^4}={s_c\over T_c^3}=
{32\over 3}\,{(y_c-1+e^{-y_c})^3\over y_c^6}e^{-y_c}{\epsilon\over T^4}\bigg\vert_{T\to\infty}.
\ee
The smallness of $L$ is due to the fact that the $y_c$-dependent term at $y_c=1.6863$ is
already considerably reduced from its maximum value $1/8$. There is no parameter with
which one could tune $L$ to a larger value.

Finally, the sound velocity is given by
\be
c_s^2={1\over 3+2y_h}\,{3-y_h-(3+2y_h)e^{-y_h}\over y_h-1+e^{-y_h}}=\fr13-{1\over(3\pi\hat T)^2}+..
\ee
At $T_c$ $c_s^2=0.024$. Note that the Chamblin-Reall background used in \cite{springer}
has a constant sound velocity and does not have a phase
transition, since for it always $\epsilon={\rm constant}\times p$.

\begin{figure}[!t]
\begin{center}

\vspace{0.5cm}
\includegraphics[width=0.49\textwidth]{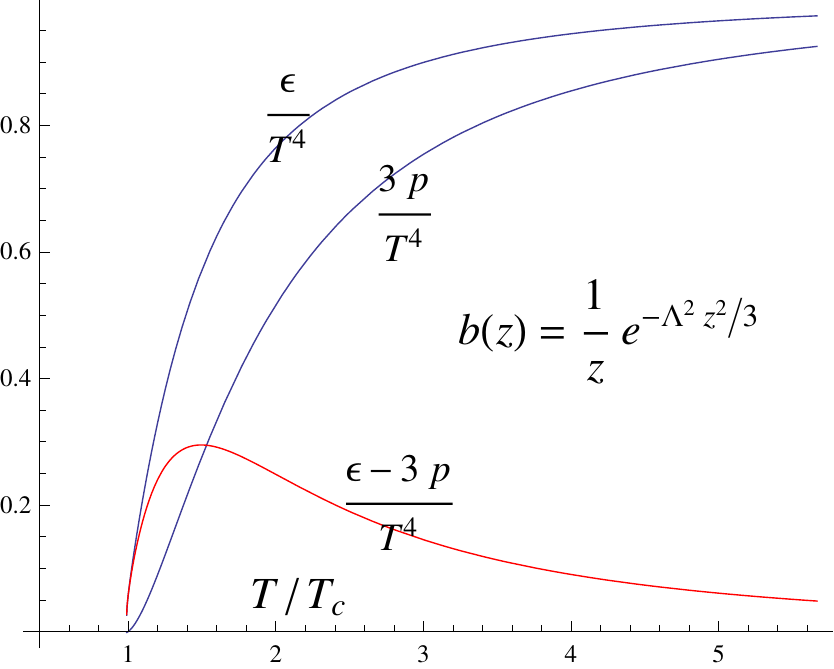}\hfill
\includegraphics[width=0.49\textwidth]{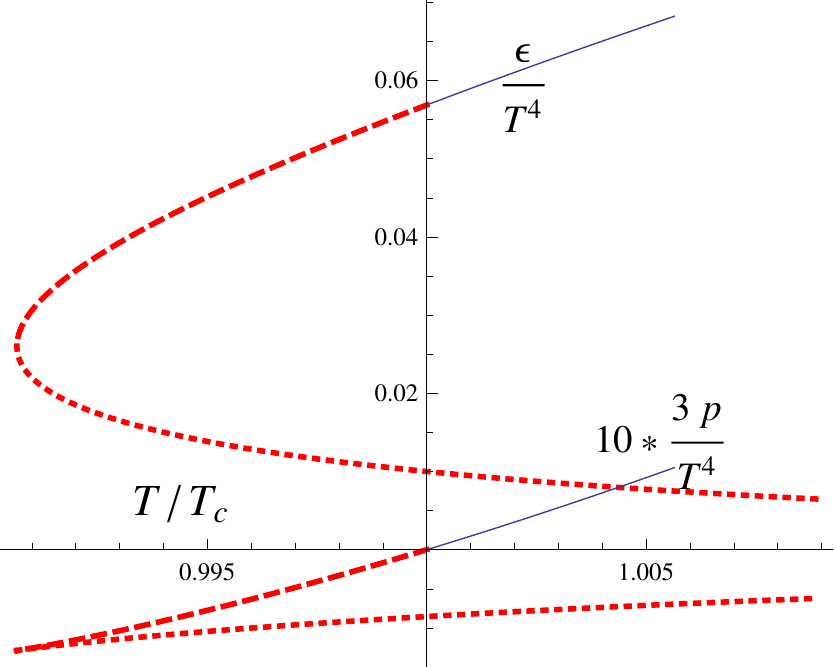}\hfill

\end{center}

\caption{\small Thermodynamics from the model $b(z)$. The curves are normalised
to their $T\to\infty$ asymptotic values, which are $\sim N_c^2$. At $T_c$ there is a
weak first order transition with $\epsilon_c/T_c^4=0.06$.
The right panel shows the region around $T_c$ with the supercooled
(dashed) and unstable (dotted) phases. The curve for pressure is multiplied by $10$ for
clarity.}
\la{modelth}
\end{figure}

\subsubsection{Glueball masses}
Glueball masses are obtained as poles of Green's function and we shall later see
how this goes through. Effectively, this leads to a Schr\"odinger-type
equation,
\be
-\psi''(z)+V_\rmi{Schr}(z)\psi(z)=m^2\psi(z),
\la{Seq}
\ee
where, for the background \nr{bg} and for scalar glueballs,
\be
V_\rmi{Schr}=\fr32\left({\ddot b\over b}+{\dot b^2\over2b^2}\right)+{\ddot X\over X}+
3{\dot b\over b}{\dot X\over X},\quad X\equiv {\beta(\lambda)\over3\lambda};
\ee
for tensor glueballs the $X-$terms are missing.
With the above explicit expressions it is trivial to evaluate
the potentials. Scaling out $\Lambda$ by writing $V/\Lambda^2\to V,\,\,\Lambda z\to z$,
the tensor potential is
\be
V_t(z)={15\over4 z^2}+2+z^2,
\ee
and the scalar one
\be
V_s(z)=V_t(z)+{1\over(3+2z^2)^2(9+2z^2)^2}\biggl(-{2187\over z^2}-5346-1224z^2+216z^4+48z^6\biggr).
\la{scalpotmodel}
\ee
The tensor equation is analytically solvable \cite{son_linconf} and leads to the mass spectrum 
\be
m_T^2/\Lambda^2=4(n+2),\quad n=0,1,2,...
\la{tensormasses}
\ee
Solving numerically the Schr\"odinger equation, one finds the
scalar glueball mass spectrum
\be
m_S^2/\Lambda^2 = 6.27,\,\,10.27,\,\,14.25,...
\ee
The lowest tensor and scalar states are in the ratio $m_{T1}/m_{S1}=1.13$
(in the full IHQCD this ratio is 1.36 \cite{kiri4}) , but the $m^2$ spacing
is about $4\Lambda^2$ also for scalars. In the full IHQCD the ratio $m_{S2}/m_{S1}=1.61$
while here it is $1.28$.

\section{Fluctuation equation}
To calculate correlators \cite{policastrostarinets,ss,kovtunstarinets,
kovtunstarinets2}
of the operator $T_{12}$ in the background \nr{bg} one has to solve
the fluctuation equation
\be
\ddot h+{d\over dz}\log(b^3f)\dot h+
\biggl({\omega^2\over f^2}-{\bfk^2\over f}\biggr)h\,\,\equiv \,\,\ddot h+P\dot h+Qh
\,=\,0.
\la{scaleq}
\ee
Here $h\equiv h_K(z)$, $K=(\omega,\bfk)$, and
\be
h(x,z)=\int{d^4K\over(2\pi)^4}e^{-i\omega t+i\bfk\cdot\bfx}h_0(K)h_K(z),\quad h_K(0)=1,
\ee
is the fluctuation of
the $12$ component of the background metric: $g_{12}=b^2(z)(1+h(t,x^3,z))$. This tensor
fluctuation does not mix with other ones. In IHQCD, the scalar one does and the fluctuation
equation permitting computations of the Green's function of the scalar operator $F_{\mu\nu}^2$
will contain more terms \cite{kiri_gravfluct,springer1,springer2}.

The characteristic exponents of \nr{scaleq} at $z=z_h$ are $\pm i\omega/\dot f_h$. Eq. \nr{scaleq}
should be solved so that at the horizon
(infalling boundary condition, $d_i$ are calculable coefficients)
\be
h_K(z\to z_h)= (z-z_h)^{i\omega/\dot f_h}[1+d_1(z-z_h)+d_2(z-z_h)^2+...]
\la{horexp}
\ee
and that at the boundary, $h_K(0)=1$ (correct normalisation). These solutions will
satisfy
\be
h_{-K}(z)=h^*_K(z).
\la{reality}
\ee

Given a solution, a computation along the lines of \cite{bls} shows that
the retarded Green's function is obtained from
\be
G(K)={1\over16\pi G_5}\biggl[fb^3\,\dot h_K h_{-K}+
({\rm counter\,\,terms})\times h_K h_{-K}\biggr],
\la{res}
\ee
where the $h_K h_{-K}=h_Kh^*_K$ terms with unspecified coefficient are real.

For the imaginary part of $G$, the spectral function, the situation is
particularly simple:
\be
{\rm Im}\,G(k)\equiv\rho(\omega,k)={1\over16\pi G_5}\,fb^3\,{\rm Im}\,\dot h_K h_{-K}=
{1\over16\pi G_5}\,fb^3\,W( h_K, h_K^*)/(2i),
\la{imG}
\ee
where one has noted that $h_K$ and $h_K^*$ are two independent solutions of \nr{scaleq}
and that their Wronskian arises when $\im G$ is evaluated. However,
the Wronskian of any two linearly independent solutions of \nr{scaleq} is,
integrating $\dot W/W=-P$,
\be
W(\phi_1,\phi_2)\equiv\phi_1\dot\phi_2-\phi_2\dot\phi_1={W_0(K)\over b^3f},
\la{Wronsk}
\ee
where $W_0$ is a $z$ independent constant (but will depend on $K=\omega,k$).
Thus the result in \nr{imG} is a ratio of two Wronskians and independent of $z$.

To find the constant $W_0$ for the Wronskian $W( h_K, h_K^*)$ we simply evaluate it
in the limit $z\to z_h$ using the expansion \nr{horexp} and find
\be
W(h_K,h_K^*)={2i\omega\over fb^3}b^3(z_h).
\ee
The result thus automatically contains the entropy density, which in IHQCD is
\be
s(T)={A\over 4G_5}={1\over4G_5}b^3(z_h).
\ee
However, so far this is for a solution normalised as in \nr{horexp},
which leads to some value $h_K(0)$
at the boundary. Correcting for the normalisation, the result is
\be
\rho(\omega,k)={1\over4\pi}s(T){\omega\over |h_K(0)|^2}.
\la{resnorm}
\ee

A practical note on the application of this formula: in it $h_K(z)$ arises from numerical
integration which cannot be extended to $z=0$ since this is a singular point of the
equation. However, in \nr{resnorm} $h_K(z)$ has to be evaluated precisely at $z=0$
and this value has to be found by extrapolation. Numerical effect is significant
at large $\omega$, where $|h_K(0)|\sim 1/\omega^{3/2}$.

We shall also need the real part, for example, the static Green's function
$G(0,k)$ is purely real.
For the real part one has to make further assumptions, corresponding to subtractions
in the dispersion relation \nr{G}. The standard procedure here \cite{kovtunstarinets}
is to expand the
solution $h_K(z)$ computed with the boundary condition \nr{horexp} in terms of the
two independent solutions at $z=0$:
\be
h_K(z)=A(K)\phi_u(z,K)+B(K)\phi_n(z,K),
\la{hAB}
\ee
where
\ba
\phi_n&=&z^4(1+c_1z^2+c_4z^4+....),\nn
\phi_u&=&1+f_1z^2+f_3z^6+...+c\log z\,\phi_n(z,K).
\la{phiexps}
\ea
In IHQCD the expansion coefficients will contain logarithmic UV coefficients of the type
in Appendix A \cite{kiri1}. Noting that the normalisation factor $h_K(0)$ in \nr{resnorm}
is $A(K)$ the result \nr{res} can be evaluated using
\ba
{h_K^*\dot h_K\over z^3|h_K(0)|^2}&=&{1\over z^3|A|^2}(A^*\phi_u+B^*\phi_n)
(A\dot\phi_u+B\dot\phi_n)\nn
&=&{2f_1\over z^2}+4c\log z+c+2f_1^2+4{B(K)\over A(K)}+\CO(z^2)\nn
&=&{\omega^2-k^2\over2z^2}-{(\omega^2-k^2)^2\over16}(4\log z-1)+4 {B(K)\over A(K)}+\CO(z^2),
\la{keyeq}
\ea
where the last form applies for conformal expansions \nr{phiexps}.
The scheme is to neglect the real divergent terms + constant in \nr{keyeq} and to evaluate
the entire Green's function from
\be
G(K)={\CL^3\over4\pi G_5}{B(K)\over A(K)},\quad {B(K)\over A(K)}=-{W(h_K,\phi_u)\over W(h_K,\phi_n)}=-
{h_K\dot\phi_u-\phi_u\dot h_K\over h_K\dot\phi_n-\phi_n\dot h_K}.
\la{Gfin}
\ee
Equivalently, one may say that this is a way to derive the counter terms in \nr{res} \cite{papa}.
In \cite{kv} this scheme was tested by evaluating the purely real static Green's function
$G(0,k)$ also by an $\omega$-integral over the complex $G(\omega,k)$. It also gave correctly the
real and imaginary parts on the light cone, for $G(k,k)$.

Since $G(K)$ in \nr{Gfin} is given as a ratio of two Wronskians, it is independent at what $z$ it is
evaluated, to the extent $h_K,\,\phi_u,\,\phi_n$ are correct solutions. Since $h_K$ is
integrated numerically, it certainly is a solution. In practice the accuracy and constancy
then depends on how many terms are included in small-$z$ expansions of $\phi_u,\phi_n$. If the
numerical integration of $h_K$ from \nr{scaleq} is extended down to some small $\epsilon$, the
ratio of Wronskians can be evaluated at this~$\epsilon$. This is to be contrasted with
\nr{resnorm} in which $h_K(0)$ has to be extrapolated to $z=0$ for accuracy. Of course, \nr{resnorm}
only gives the imaginary part while both the real and imaginary parts are obtained from \nr{Gfin}.

\subsection{Equations for the model}
Next, we specialize to the model defined in subsection \ref{subsectmodel}. In terms of
$y=\Lambda^2z^2$ the fluctuation equation \nr{scaleq} becomes
\be
\phi_K''(y)+P(y)\phi_K'(y)+Q(y)\phi_K(y)=0.
\la{PQ}
\ee
where
\be
P(y)={d\over dy}\log( b^3f\sqrt{y})=-{1\over y}-1+{y \over y-1+(1-y_h)e^{y_h-y}},
\ee
\be
Q(y)={\hat\omega^2\over y f^2(y)}-{\hat k^2\over y f(y)},
\ee
where the dimensionless $\hat\omega,\hat k$ were defined in \nr{enscales}.

The equation \nr{PQ} is to be solved so that at the horizon (the $y-y_h$ term is for $k=0$)
\be
\phi_K(y)=(y-y_h)^p\biggl[1+{p(1+y_h)-2p(2+y_h)\over2y_h(1+2p)}(y-y_h)+\CO((y-y_h)^2)\biggr].
\ee
where
\be
p={-i\omega\over4\pi T}=i{\hat\omega\over\sqrt{y_h}f'(y_h)}=
-i{\hat\omega\over y_h^{3/2}}(y_h-1+e^{-y_h}).
\ee
For correct normalisation, the solution is divided by $\phi_K(0)$.

The Green's function now is
\be
G(K)={2\Lambda\over16\pi G_5}\biggl[fb^3\sqrt{y}\,\phi'_K\phi_{-K}+
{\rm counter\,\,terms}\,\,\cdot\phi_K\phi_{-K}\biggr].
\la{Gy}
\ee
For the spectral function one again needs the Wronskian of the two
independent solutions at the horizon,
\be
W(\phi_K,\phi_K^*)={2i\hat\omega\over fb^3\sqrt{y}}b^3(y_h),
\ee
and the result \nr{resnorm} is unchanged. For the real part and evaluation of the
counter terms in \nr{Gy} the expansions around $y=0$ have to be worked out.
One writes
\be
\phi_K(y)=A(K)\phi_u(y,K)+B(K)\phi_n(y,K),
\la{phiy}
\ee
where
\be
\phi_n(y,K)=y^2\biggl[1+\fr13(2+k^2-\hat\omega^2)y+...\biggr],
\ee
\be
\phi_u(y,K)=1+(\hat\omega^2-k^2)y-
\fr12[(\hat\omega^2-\hat k^2)^2-(\hat\omega^2-\hat k^2)]\log y\cdot \phi_n(y)+\CO(y^3)+...,
\ee
(note no $y^2$ term; it is only in $\phi_n$). The real counter terms neglected now are
\be
2\Lambda^4\biggl\{{\hat\omega^2-\hat k^2\over y}-
\fr12 [(\hat\omega^2-\hat k^2)^2-(\hat\omega^2-\hat k^2)]\,(2\log y+1)\biggr\}
\ee
and, solving $A$ and $B$ with Wronskians,
\be
G(K)={\CL^3\over4\pi G_5}\cdot \Lambda^4\,{B(K)\over A(K)}.
\la{GyBA}
\ee
Now the correct dimension 4 comes from $\Lambda^4$. In units of $\pi T$, using \nr{tyh}:
\be
{G(K)\over(\pi T)^4}={\CL^3\over4\pi G_5}\cdot
{16(y_h-1+e^{-y_h})^4\over y_h^6}\cdot{B(K)\over A(K)}.
\la{GinpiT}
\ee
We are interested in the high $T$ phase and, as was derived earlier,
this means $T\ge T_c=0.4000\Lambda$ or $y_h<y_c=1.6863$.

\subsection{Vacuum spectral function and the limit at large $\omega$ or $k$}\la{largeK}
To obtain the vacuum Green's function of the model we put $f=1$ in \nr{PQ} and obtain the
equation
\be
y\phi''(y)+(-1-y)\phi'(y)+ \hat K^2\phi(y)=0,\quad \hat K^2\equiv\hat\omega^2-\hat k^2.
\la{diffeq}
\ee
This also gives the leading terms of the $\omega,k\gg\Lambda,T$ limit 
of the solutions of \nr{PQ}.
Understanding this limit analytically is important for numerical accuracy.
This equation is the equation for the confluent hypergeometric function with a solution
$\phi_1$ growing like a power at large $y$:
\be
\phi_1(y)=y^2\,U(2-\hat K^2,3,y)
\to y^{\hat K^2}\bigl[1+\hat K^2(2-\hat K^2)y^{-1}+\CO(y^{-2})\bigr].
\ee
Its small $y$ expansion is
\be
\phi_1(y)={1\over\Gamma(2-\hat K^2)}\biggl[1+\hat K^2y+\fr12 \hat K^2(1-\hat K^2)
\biggl(\log y+\psi(2-\hat K^2)+2\gamma_E-\fr32\biggr)y^2+\CO(y^3)\biggr].\la{smally}
\ee
The second solution $\phi_2(y)=y^2L_{\hat K^2-2}^{(2)}(y)\to e^y\,y^{-\hat K^2+1}+..$,
grows exponentially and is unacceptable. If we now decompose the
solution $\phi_1$ as in \nr{phiy}, $B$ is immediately read as the coefficient
of the $y^2$ term and $A=1/\Gamma(2-\hat K^2)$. The vacuum (this is vacuum since
no $T$ is involved) Green's function then is
\be
G\rmi{vac}(K)=\Lambda^4\cdot \fr12 \hat K^2(1-\hat K^2)
\biggl[\psi(2-\hat K^2)+2\gamma_E-\fr32\biggr].
\la{gas}
\ee
Since
\be
\psi(z)=-\gamma_E+\sum_{m=0}^\infty\biggl[{1\over m+1}-{1\over m+z}\biggr],
\ee
the result in \nr{gas} has poles at
\be
\omega^2-k^2=4\Lambda^2(m+2).
\ee
This reproduces the tensor glueball mass spectrum in \nr{tensormasses}, now derived as
poles of Green's function. Taking $\omega\to\omega+i\epsilon$ and plotting $\im G_\rmi{vac}$
from \nr{gas} for finite but small $\epsilon$ one finds Gaussian peaks at glueball masses.
In the limit $\epsilon\to0$ \cite{forkel}, 
\be
\rho_\rmi{vac}\equiv\im G_\rmi{vac}=
\fr\pi{32}(\omega^2-k^2)(\omega^2-k^2-4\Lambda^2)
\sum_{m=0}^\infty\,\delta\biggl(m+2-{\omega^2-k^2\over4\Lambda^2}\biggr).
\la{imgas}
\ee
We shall later use this as a baseline to assess the magnitude of finite $T$
effects (see Eq.~\nr{Gmiddlevac}).

The above was for vacuum, $f=1$. The asymptotic large $\omega,k$ limit for $f\not=1$ can be argued
as follows. Taking first $\omega>k>0$, the large $\hat K^2>0$ limit of $\psi$ in 
\nr{gas} is
\be
\psi(2-\hat K^2)=\pi \cot(\pi\hat K^2)+\log\hat K^2-{3\over\hat 2K^2}-{13\over12\hat K^4}+...\,.
\la{psilim}
\ee
Note now that $\pi\cot(x+iy)\to\mp i\pi$ if $y\to\pm\infty$. If now the analytic continuation
is performed so that also $\im\,\hat K^2\to\infty$, then \nr{gas} and \nr{psilim} give
\be
{\im G(K)\over\Lambda^4}={\pi\over2}\left[(\hat\omega^2-\hat k^2)^2-(\hat\omega^2
-\hat k^2)+\CO(1)\right].
\la{imgasfull}
\ee
Both terms are very clearly seen in numerics, but we have been unable to fit a constant
(or possibly a logarithmic) term nor compute it analytically.

In contrast, in the static $\omega=0$ case, $\hat K^2=-\hat k^2$, the constant term
(and higher corrections $\CO(1/k^2)$) can be evaluated. Now
Eq.~\nr{gas} is purely real.
Expanding $\psi(2+k^2)\to \log k^2+3/(2k^2)-13/(12 k^4)+\CO(1/k^6)$ in \nr{gas} gives
a constant term $-5/24$. However, this is not the entire
constant term, which is crucial for numerical analysis.
A careful analysis (see Appendix \ref{app:olver}) permits one to compute it with the result
\nr{careful}, which is the same as writing \nr{gas} in the form
\be
{B\over A}\to-\fr12\hat k^2(1+\hat k^2)\left[\psi(2+\hat k^2)+2\gamma_E-\fr32\right]
+{e^{-y_h}\over20(y_h-1+e^{-y_h})}.
\la{staticvac}
\ee
Thus there is an overall constant term
$-5/24+e^{-y_h}/(20(y_h-1+e^{-y_h}))\approx -5/24+T_c^4/(5T)^4$.
This is clearly seen in
numerics.

In the thermal theory with $f\not=1$ the glueballs form a glueball gas
with pressure $\CO(N_c^0)$
and AdS/QCD duality says little about its properties \cite{noronha},
it is the phase at $p=0$ in Fig.~\ref{modelth}. In the high $T$ phase at $T\ge T_\rmi{min}$
the delta function peaks are not seen but still
some structure remains (see Figs.~\ref{rhok0} and \ref{k0mG} below).
The broadening of hadronic states in a thermal ensemble has been studied
extensively in connection with quarkonium physics \cite{mocsy}. 

\begin{figure}[!tb]
\begin{center}

\vspace{0.5cm}
\includegraphics[width=0.49\textwidth]{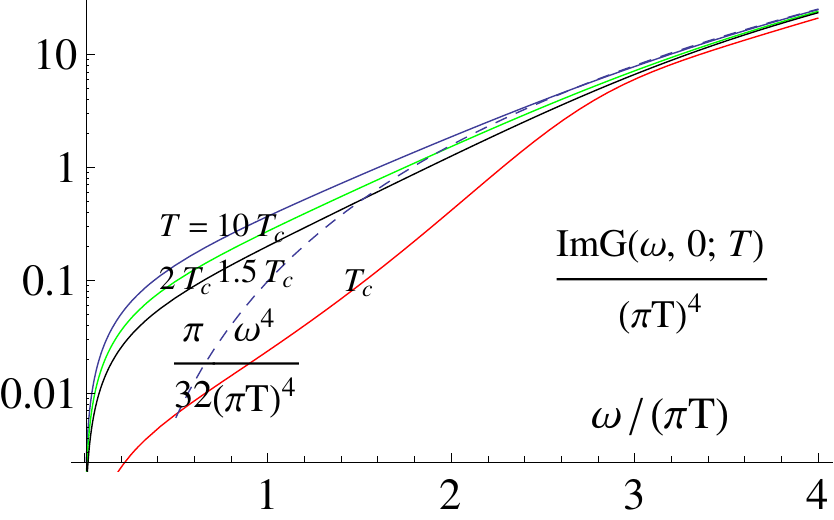}\hfill
\includegraphics[width=0.49\textwidth]{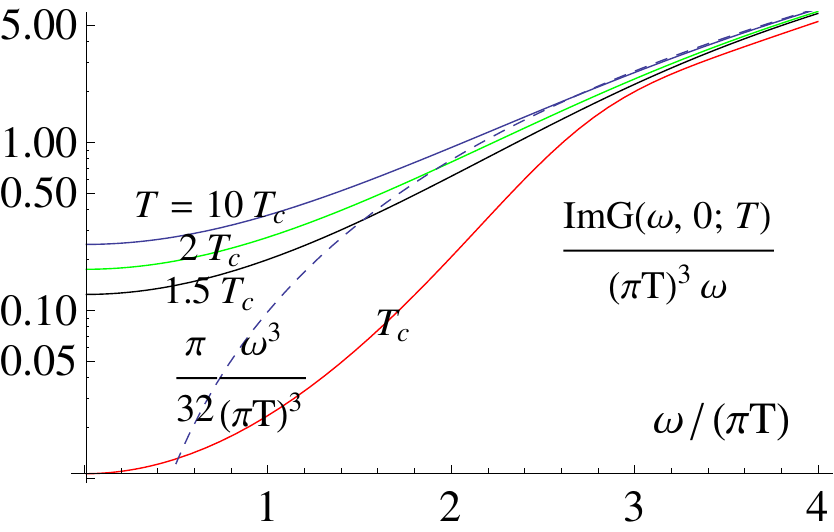}

\end{center}

\caption{\small The spectral function $\im G(\omega,0;T)/(\pi T)^4$ in units of $\CL^3/(4\pi G_5)$
plotted vs $\omega/(\pi T)$ (left panel) or divided by $\omega/(\pi T)$ (right panel), for the
model \nr{bmodel}. The dashed line is the leading large-$\omega$ limit.}
\la{spectfn}
\end{figure}

\begin{figure}[!tb]
\begin{center}

\vspace{0.5cm}
\includegraphics[width=0.49\textwidth]{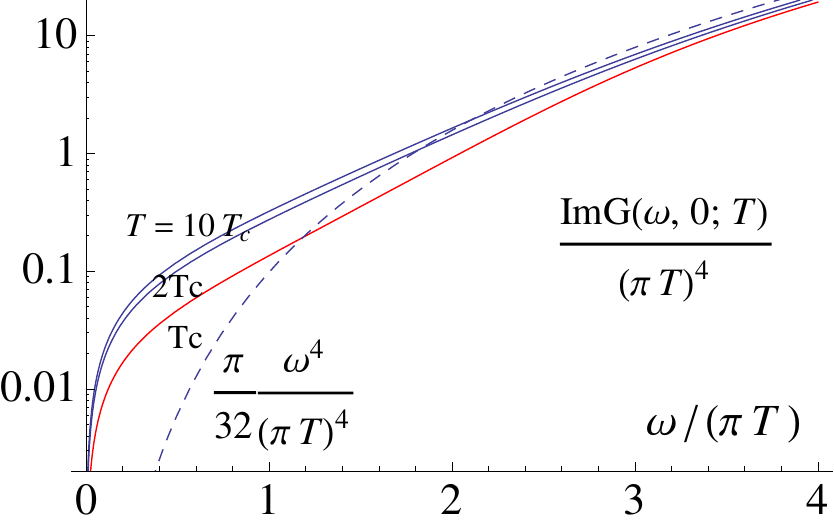}\hfill
\includegraphics[width=0.49\textwidth]{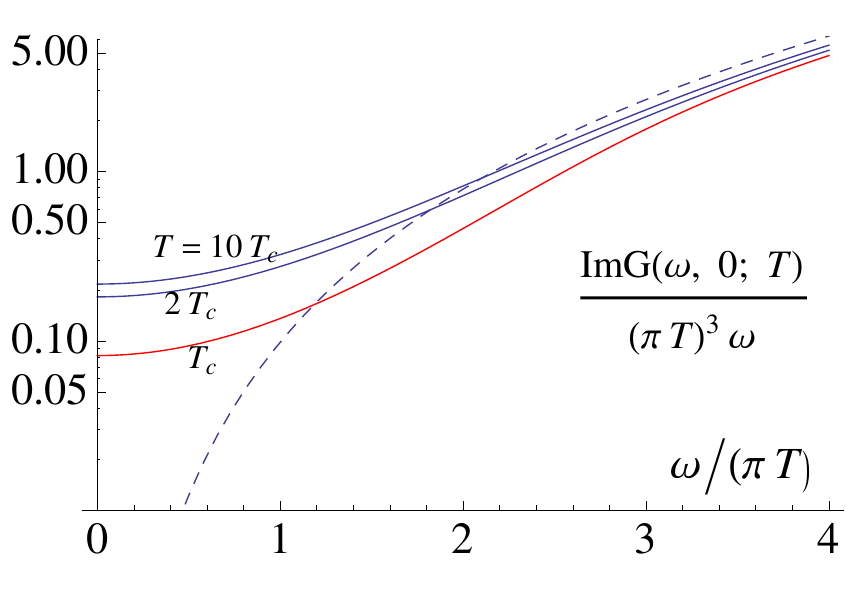}

\end{center}

\caption{\small The spectral function $\im G(\omega,0;T)/(\pi T)^4$ in units of $\CL^3/(4\pi G_5)$
plotted vs $\omega/(\pi T)$ (left panel) or divided by $\omega/(\pi T)$ (right panel) in exact numerics
of IHQCD. }
\la{rhotot}
\end{figure}

\begin{figure}[!t]
\begin{center}

\vspace{0.5cm}

\includegraphics[width=0.49\textwidth]{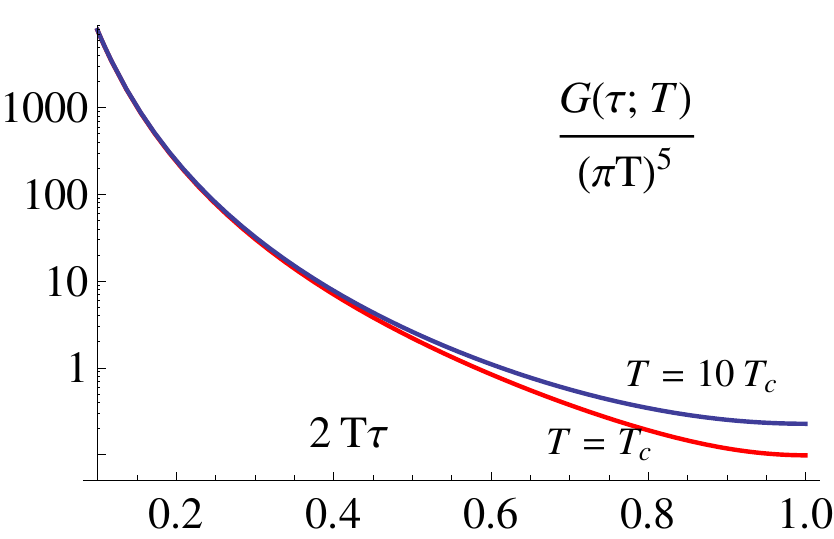}\hfill
\includegraphics[width=0.49\textwidth]{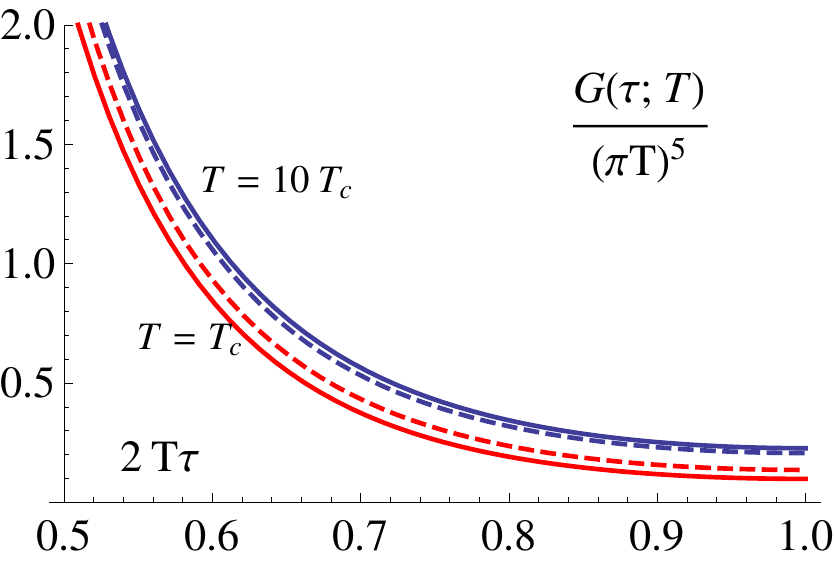}

\end{center}

\caption{\small The imaginary time Green's function $G(\tau,k=0)$ computed from \nr{gtauint}. The
right panel shows the region near $\tau=1/(2T)$ on a linear scale, the dashed curves come
from full numerical evaluation of IHQCD.}
\la{gtaumodel}
\end{figure}

\section{Numerical results}
We are interested in computing the finite temperature retarded Green's function of
$T_{12}$ in a QCD-like strongly interacting theory, for various
frequences and wave numbers $\bfk=(0,0,k)$.
At large $T$, say, $T>10T_c$ one expects the theory to approach
a conformal theory with Green's functions explicitly plotted in
\cite{kovtunstarinets2,teaney,kv}. In the range $T_c$
to $10T_c$ there are important nonconformal effects, reflected in the large value of the
interaction measure $(\epsilon-3p)/T^4$. This is the range of $T$ to be studied here.

We remind that the results for $G$ or $\rho=\im G$ are given without an overall
factor of $\CL^3/(4\pi G_5)$.
\subsection{Spectral function for $k=0$}
This is clearly a very interesting range due to its relation to viscosity and lattice
determinations thereof \cite{karschwyld,aarts,meyershear,meyerbulk,meyerlat08,meyer_review}.
Consider first the results for the model \nr{bmodel}.
As discussed in
subsection \ref{largeK}, the large-$\omega$ limit is, measuring both $G$ and $\omega$ in
units of $\pi T$,
\be
{\rho_\rmi{as}(\omega,k)\over(\pi T)^4}={\im G_\rmi{as}(\omega,k;T)\over(\pi T)^4}
={\pi\over32}\biggl[(\tilde\omega^2-\tilde k^2)^2-
{25.0T_c^2\over(\pi T)^2}(\tilde\omega^2-\tilde k^2)\biggr]\theta(\omega-k).
\la{TdepinG}
\ee
This is an asymptotic formula and both terms are clearly seen in numerics. However, we
do not know the terms of lower order in $\omega^2-k^2$, but \nr{TdepinG} should be applicable only
for $\omega^2>k^2+25.0T_c^2$, it is negative at smaller $\omega$.

Results for the model \nr{bmodel} are shown in Fig.~\ref{spectfn}
together with the leading asymptotic behavior $\pi \omega^4/32$.
The spectral function is shown both as such or divided by $\omega$ so that the value at $\omega=0$
is essentially the shear viscosity. One observes a sizeable effect near $T_c$ but already at $2T_c$
the curves are very close to the conformal situation, approximated by the $10T_c$ curve.
The main $T$-dependent effect is that contained analytically in the asymptotic behavior \nr{TdepinG}.
Even for
$T\approx T_c$ the nonconformal effects disappear at $\omega\gsim3T_c$; remember that the lightest
scalar glueball mass is $\approx6.2T_c$.

Consider then the results, shown in Fig.~\ref{rhotot}, computed from full
numerics\footnote{The parameters of the scalar potential were otherwise the same as those in \cite{kiri4}
but $V_0=1$ instead of $V_0=0.04128$; this makes numerics simpler and does not affect the results.}
of IHQCD, using the formula \nr{resnorm}. This gives only the imaginary part;
due to the complicated logarithmic structure of IHQCD (see Appendix A) it will be more
complicated to evaluate the Wronskians in \nr{Gfin} and compute the real part also.
The overall pattern at $k=0$ is very similar. An interesting detail is that in IHQCD the
approach to the asymptotic limit is rather slow, possibly related to the UV
logarithms.

The imaginary time $T_{12}$ correlator at $\bfk=0$ can now be integrated from
\be
G(\tau,0;T)=\int_0^\infty {d\omega\over\pi}\rho(\omega,0;T)
{\cosh[(1-2T\tau)\fra\pi 2\omega]\over\sinh(\fra\pi2\omega)},
\la{gtauint}
\ee
with everything in units of $\pi T$. Note first that in the $\CN=4$ conformal case
the answer is \cite{kv}
\be
G(\tau,0;T)={3\over4\pi^5}[\zeta(5,T\tau)+\zeta(5,1-T\tau)]+G_\beta(T\tau),
\la{Gtau0}
\ee
where the first terms come from the integration of $\pi\omega^4/32$ and $G_\beta$ from
$\rho(\omega,0)-\pi\omega^4/32$. The thermal part $G_\beta$ varies very slowly from
$G_\beta({\fra1{2}})=0.0710$ at the symmetry point to the value $0.13$ at $\tau=0$. 
At $T\tau={\fra12}$ the vacuum part, i.e., the $\zeta$
function terms, contribute $93\zeta(5)/(2\pi^5)=0.1576$ 
and dominate totally at smaller $\tau$.

\begin{figure}[!t]
\begin{center}

\vspace{0.5cm}

\includegraphics[width=0.49\textwidth]{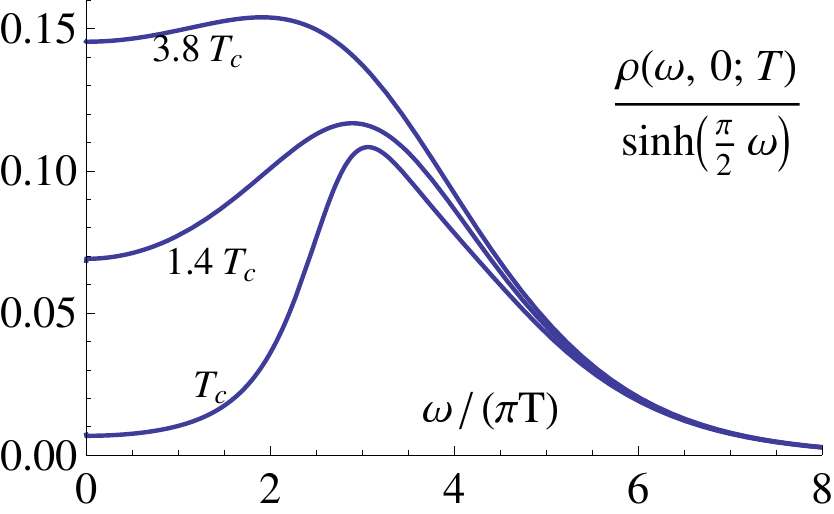}\hfill
\includegraphics[width=0.49\textwidth]{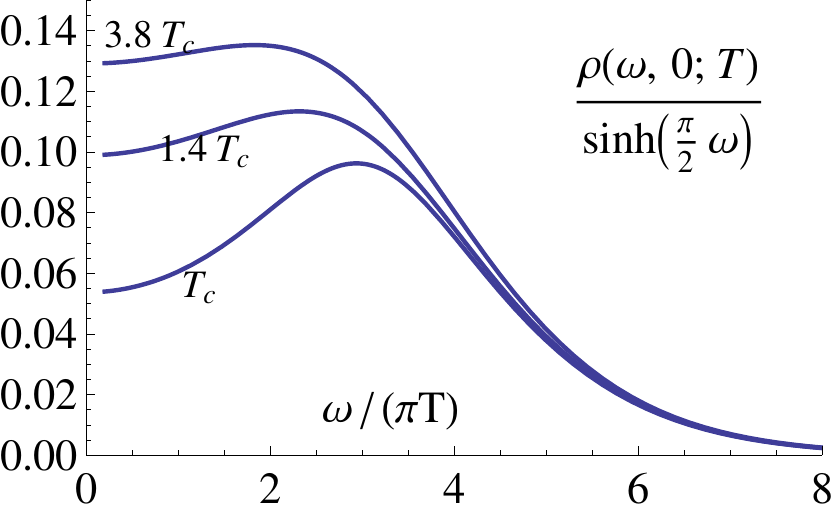}

\end{center}

\caption{\small The weighted spectral distribution, integrand of \nr{gtauint},
at the middle point $\tau=1/(2T)$ for both the model \nr{bmodel} (left) and full numerics
of IHQCD (right), at various temperatures.
}
\la{middlespect}
\end{figure}

\begin{figure}[!t]
\begin{center}

\vspace{0.5cm}

\includegraphics[width=0.49\textwidth]{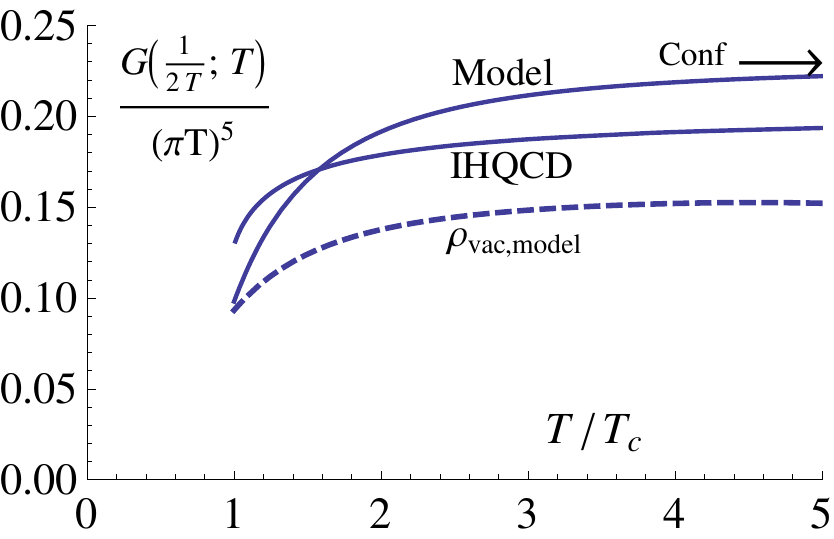}

\end{center}

\caption{\small The temperature dependence of $G(1/(2T))$ for both the model \nr{bmodel}
and full numerics of IHQCD. The arrow shows the value for the conformal $\CN=4$ theory \cite{kv}
The dashed curve shows the plot of \nr{Gmiddlevac}, generated from the $T$ independent
vacuum spectral function \nr{imgas} via the spectral representation \nr{gtauint}.
}
\la{gtaumiddle}
\end{figure}

The outcome in the nonconformal cases is shown in Fig.~\ref{gtaumodel}. The $T=10T_c$
curve is very close to the conformal curve \nr{Gtau0}.
A particularly interesting point is the middle one, $\tau=1/(2T)$. Fig.~\ref{middlespect}
shows the corresponding integrand of \nr{gtauint}, in which large $\omega$ values are strongly
suppressed, Fig.~\ref{gtaumiddle} shows the resulting value of $G(\tau=1/(2T))$.

To assess the significance of the result, it is very useful to generate a comparison
curve by inserting the vacuum spectral function \nr{imgas} with delta function peaks
at the position of glueball masses to the representation \nr{gtauint} at $\tau=1/(2T)$.
The answer is ($\Lambda=2.5T_c$)
\be
{G_\rmi{vac}(\tau=1/(2T))\over(\pi T)^5}=\fr12\biggl({2.5T_c\over\pi T}\biggr)^5
\sum_{m=0}^\infty{\sqrt{m+2}(m+1)\over\sinh(2.5\sqrt{m+2}\,T_c/T)}.
\la{Gmiddlevac}
\ee
The resulting curve is also plotted in Fig.~\ref{gtaumiddle}, the limit at large
$T$ is $93\zeta(5)/(2\pi^5)$, the same as for the vacuum curve in \nr{Gtau0}.
One sees that the imaginary time correlator
generated by the vacuum spectral function behaves similarly to the genuine finite $T$
correlator, finite $T$ just enhances the magnitude. The vacuum correlator has nothing
like a transport peak at $\omega=0$, it is identically zero up to the lowest glueball
mass at $2\sqrt2\Lambda$. Still the outcomes are very similar. 

We do not attempt a detailed comparison with lattice Monte Carlo data. However, the results
are in qualitative agreement with the plots in \cite{meyershear}. An agreement confirmed by
careful analysis would support the present picture of strongly interacting QCD matter, but 
Fig.~\ref{gtaumiddle} illustrates the difficulty of making definite conclusions.

Since one knows that the viscosity prediction holds independent of $T$ it is useful to
present the data so that this is explicitly seen. Data for the dimensionless quantity
\be
{4\pi\over s(T)}{\rho(\omega,0;T)\over\omega},
\la{rhoom}
\ee
computed for the model \nr{bmodel}
in the region of small $\omega$, are presented in Fig.~\ref{rhok0}, left panel.
The normalisation is such that the standard viscosity prediction
$\eta/s=1/(4\pi)$ corresponds to the value 1 at $\omega=0$. The right panel of
Fig.~\ref{rhok0} shows the spectral function with the asymptotic behavior
\nr{TdepinG} subtracted. The subtraction is reliable only for $\omega\gsim 5.0T_c/(\pi T)$.

Finally, Fig.~\ref{k0mG} shows the subtracted Green's function plotted in non-thermal
units, vs. $\omega/(2\Lambda)$, computed for the model \nr{bmodel}.
This is what remains of the infinite set of delta function peaks in \nr{imgas}.
One may try to define a glueball mass at finite temperature by the position
of the first peak in this curve.
This grows very accurately linearly with
$T$:
\be
\hat m_G(T)={m_G(T)\over m_\rmi{G}/\sqrt2}=1.016{T\over T_c}-0.241,
\ee
where the lowest tensor glueball mass is $m_G=2\sqrt{2}\Lambda$.

\begin{figure}[!tb]
\begin{center}

\vspace{0.5cm}
\includegraphics[width=0.49\textwidth]{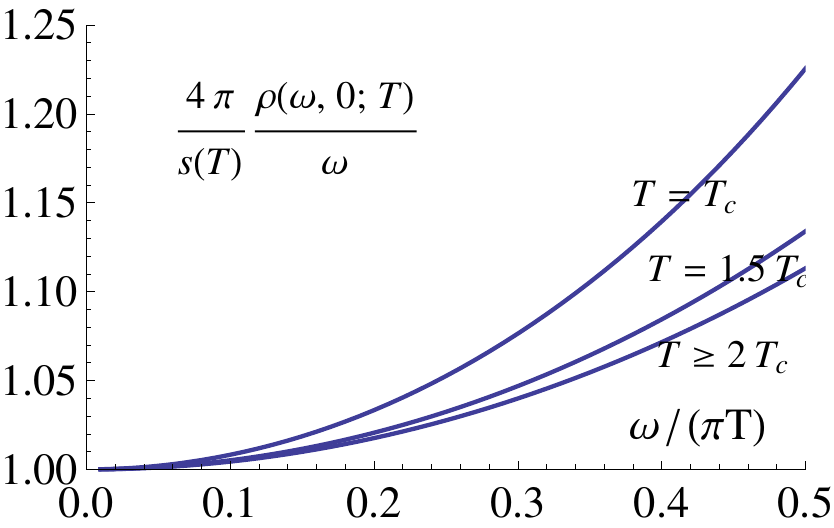}\hfill
\includegraphics[width=0.49\textwidth]{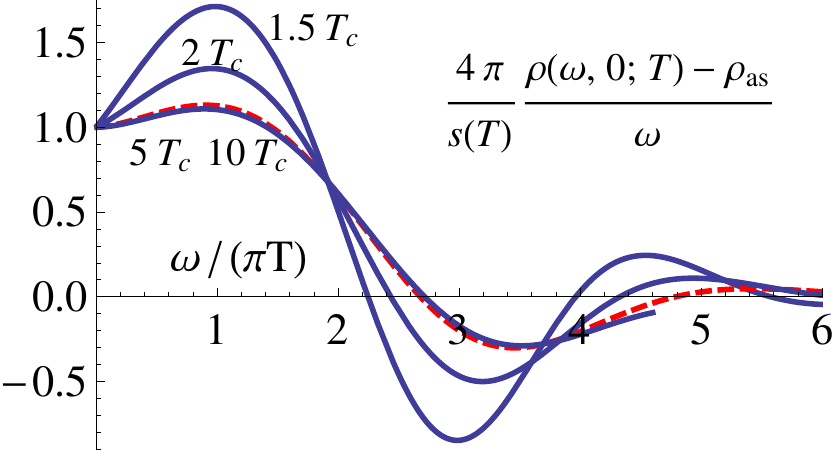}

\end{center}

\caption{\small The imaginary part of the Green's function $G(\omega,0;T)$
normalised as in \nr{rhoom}
plotted vs $\omega/(\pi T)$ for small $\omega$ (left panel) or larger $\omega$ with the
asymptotic behavior \nr{TdepinG} subtracted (right panel). The subtraction is reliable only
for $\omega/(\pi T)\gsim 5.0T_c/(\pi T)$. The model \nr{bmodel} is used here.}
\la{rhok0}
\end{figure}

\begin{figure}[!tb]
\begin{center}

\vspace{0.5cm}
\includegraphics[width=0.6\textwidth]{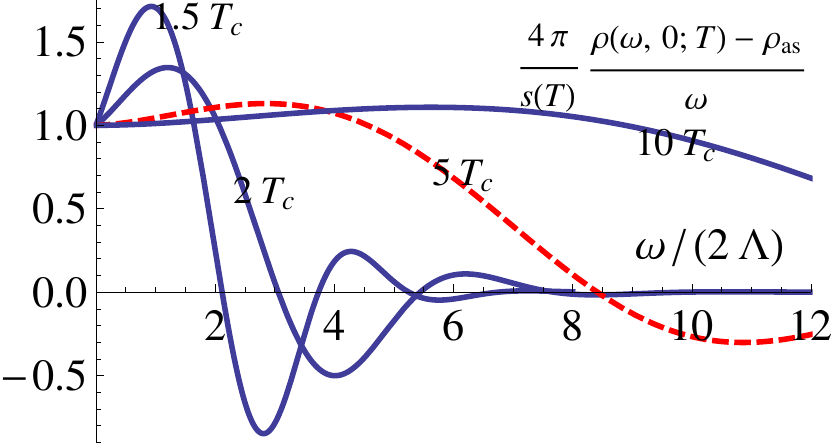}
\end{center}

\caption{\small The imaginary part of the Green's function $G(\omega,0;T)$
with the asymptotic behavior subtracted, normalised as in \nr{rhoom}
plotted vs $\hat \omega=\omega/(2\Lambda)$. The model \nr{bmodel} is used here.}
\la{k0mG}
\end{figure}

\subsection{Real static correlator, $\omega=0$}
The static correlator is defined by
\be
G(r)=\int_0^\beta d\tau\,\langle  T_{12}(\tau,\bfx)T_{12}(0,{\bf 0})\rangle_T=
\int_0^\beta d\tau \,G(\tau,\bfx)={1\over2\pi^2 r}\int_0^\infty dk\,k \sin(r k)G(k)
\la{gk}
\ee
and can also be related to the $\omega$-dependent spectral function by
\be
G(k)=\int_0^\infty{d\omega\over\pi}{2\rho(\omega,k)\over\omega}.
\la{gkomega}
\ee
It is determined numerically by solving \nr{scaleq} with $\omega=0$ so that the boundary
condition at the horizon is \nr{horexp} with $\omega=0$ and solving $G(0,k)$ from
\nr{Gfin} as a ratio of two Wronskians. It is also important to eliminate the
large-$k$ behavior given in \nr{staticvac}, including the constant term. Upon
transforming to coordinate space these give rise to $\delta(\bfx)$ or its derivatives.
To get $G(0,k;T)/T^4$ one further has to use \nr{GinpiT}.

\begin{figure}[!t]
\begin{center}

\vspace{0.5cm}
\includegraphics[width=0.49\textwidth]{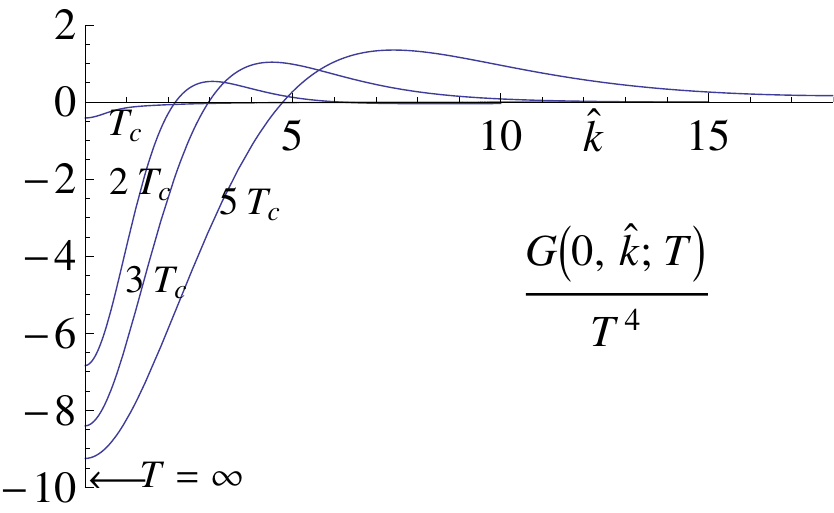}\hfill
\includegraphics[width=0.49\textwidth]{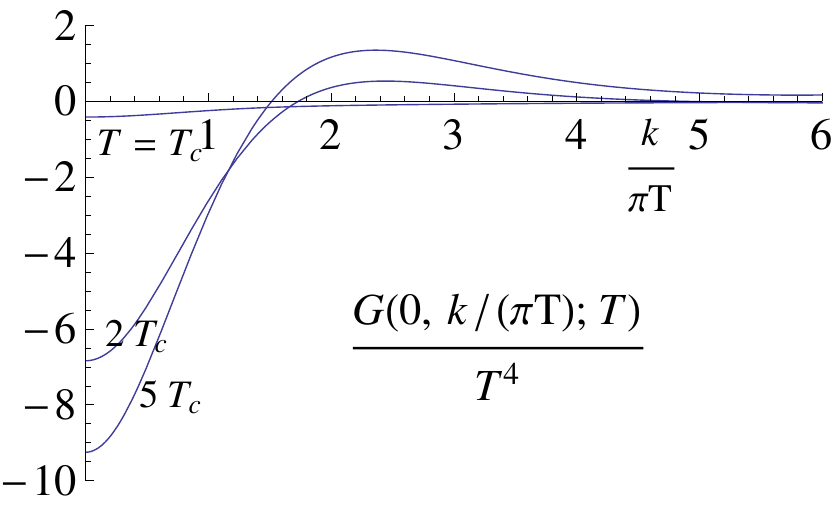}

\end{center}

\caption{\small The (real) static Green's function $G(0,k;T)/T^4$ in units of $\CL^3/(4\pi G_5)$
with the asymptotic behavior \nr{staticvac} subtracted
plotted vs $\hat k=k/(2\Lambda)$ (left panel) or vs $k/(\pi T)$ (right panel). }
\la{statcorr}
\end{figure}

\begin{figure}[!tb]
\begin{center}

\vspace{0.5cm}
\includegraphics[width=0.4\textwidth]{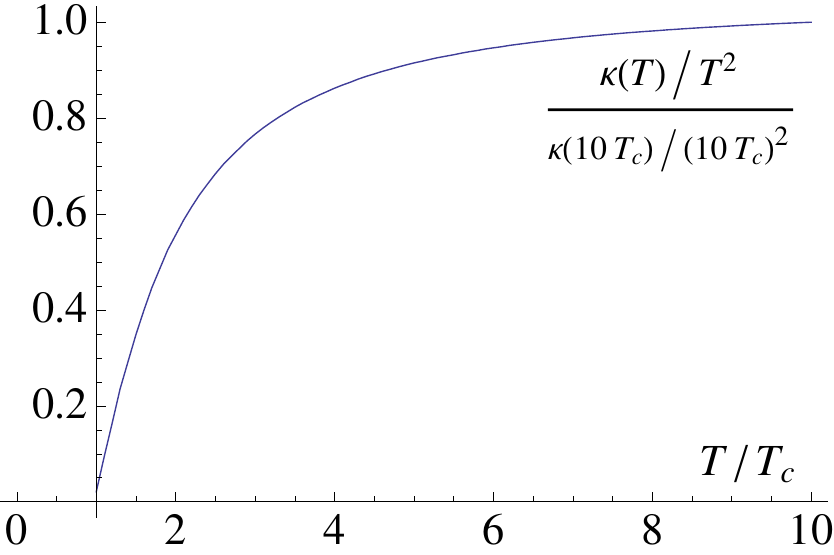}

\end{center}

\caption{\small The expansion coefficient $\kappa(T)$ defined by \nr{kappaT}, normalised
to the value of $\kappa(T)/T^2$ at $10T_c$.
}
\la{kappaT}
\end{figure}

Subtracting the expansion \nr{staticvac} from the numerically computed $B/A$
gives the result in Fig.~\ref{statcorr} for the static correlator.
The curves are of the same shape as those in the conformal case \cite{kv}, where one
plots vs $k/(\pi T)$. When $T\to\infty$ the value at $k=0$ of the curves approach the
value $-\pi^4/10$ of the conformal case. The dominant scale breaking effect is the
rapid decrease of the correlator when $T$ decreases towards $T_c$. The next subsection
explains how this is via Eq.~\nr{gkomega} related to the behavior of $\rho(\omega,k)$ 
near the light cone, see Fig.~\ref{crossLC}.

In the conformal case one can define a second order dissipative coefficient by
the expansion
\be
G(0,k;T)= G_0+{\fra12}\kappa(T) k^2+\CO(k^4)
\la{defkappa}
\ee
and its value is \cite{baier}
\be
\kappa ={\CL^3\over4\pi G_5}{\fra14}\pi^2T^2={\fra18}N_c^2T^2= {s\over4\pi^2 T}.
\la{kappa}
\ee
Inserting the normalisation \nr{Lthree}, our computation of $\kappa$
at $10T_c$ is 1\% from the conformal value in \nr{kappa}. At smaller $T$
there is additional suppression on top of the $T^2$ dependence. This is
quantitatively shown in Fig.~\ref{kappaT}.

Given the momentum dependent static Green's function $G(k)$, the coordinate
dependent correlator $G(r)$ can be computed by carrying out the integral \nr{gk}.
The shape of the curves is very similar to that plotted explicitly for the
conformal case in \cite{kv}.

\subsection{Green's function for any $\omega,k$}
The dominant feature on the $\omega,k$ plane is the structure around the light cone.
In \cite{kv} this was analysed analytically and, for example, one could prove that
exactly on the light cone
\be
G(k,k)= (1+i\sqrt3){\Gamma(1/3)\over8\cdot6^{1/3}\Gamma(2/3)}\,k^{4/3}
+\CO(1)
\approx(0.136092+i\,0.235718)\,k^{4/3},\quad k\equiv k/(\pi T).
\la{LCresult}
\ee

Results for the spectral function as a function of $\omega$ at fixed values of $k$
are shown in Fig.~\ref{k02}. The left panel shows what effects changing $T$ at fixed value
of $k/(\pi T)$ has: the correlation function monotonically decreases. The right panel
shows that essentially the conformal set of curves is obtained for $T=10T_c$, for
the model \nr{bmodel}.

Another useful way of analysing the light cone region is to see how the spectral function
varies when crossing the light cone perpendicular to it, i.e. at fixed $\omega_+=
(\omega+k)/\sqrt2$. The outcome is shown in Fig.~\ref{crossLC} at a fixed value of
$\omega_+$. At larger $\omega_+$ the peak gets narrower $\sim1/k^{1/3}$ but higher
$\sim k^{1/3}$ so that the integral over $\omega$, which via Eq.~\nr{gkomega} leads
to the static correlator, is essentially constant. This is as shown in \cite{kv} for
the conformal case. The new element is that when $T$ decreases towards $T_c$, the
correlator also decreases. After integration over $\omega$ according to \nr{gkomega}
this correlates with the rapid decrease of the static correlator, already shown in
Fig.~\ref{statcorr}.

\begin{figure}[!tb]
\begin{center}

\vspace{0.5cm}
\includegraphics[width=0.49\textwidth]{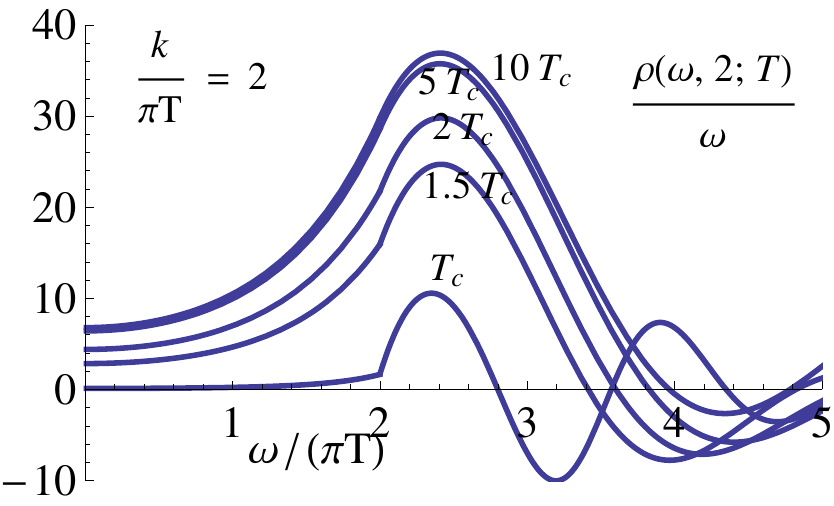}\hfill
\includegraphics[width=0.49\textwidth]{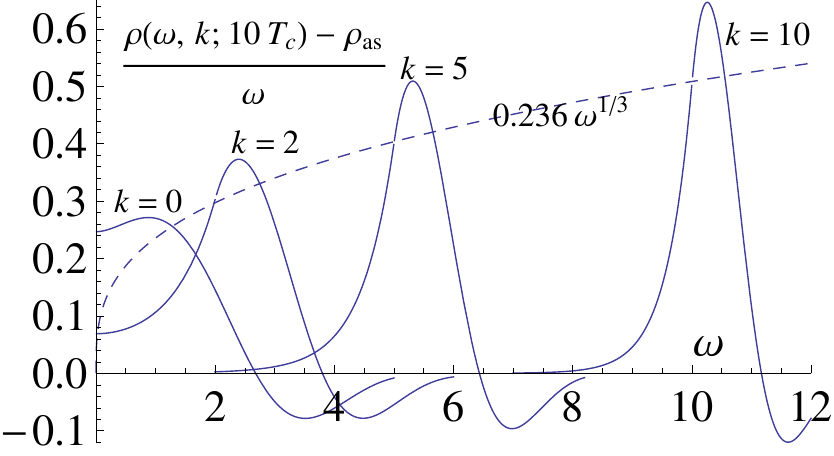}

\end{center}

\caption{\small The imaginary part of the Green's function $G(\omega,k=2\pi T;T)/T^4$
in units of $\CL^3/(4\pi G_5)$ plotted vs $\omega/(\pi T)$, for the model \nr{bmodel}.
The right panel shows the behavior
near the peaks at $k=2,5,10\pi T$. The curve $0.236\omega^{1/3}$ is the conformal behavior
of the correlator on the light cone, $\omega=k$. }
\la{k02}
\end{figure}

\begin{figure}[!tb]
\begin{center}

\vspace{0.5cm}
\includegraphics[width=0.49\textwidth]{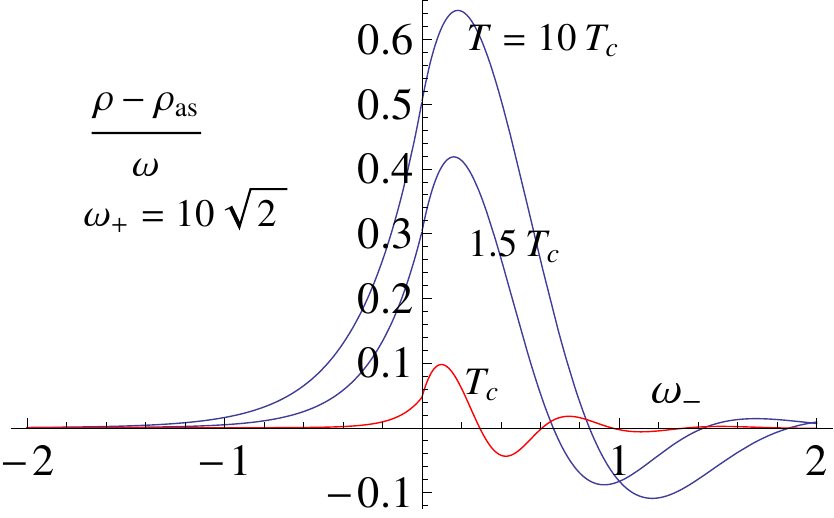}
\end{center}

\caption{\small The quantity $(\rho(\omega,k;T)-\rho_\rmi{as}(\omega,k))/\omega$ when crossing
the light cone in the perpendicular direction, i.e., at fixed $\omega_+=(\omega+k)/\sqrt2$ (here
$\omega_+=10\sqrt2$) with varying $\omega_-$, for various temperatures.}
\la{crossLC}
\end{figure}

\section{Conclusions}
We have in this paper computed the retarded Green's function $G(\omega,k;T)$,
$\bfk=(0,0,k)$, of the $1,2$ component of the energy momentum tensor in
the high temperature phase of a QCD-like
strongly coupled theory with a first order phase transition and glueballs,
using gauge/gravity duality. Temperatures between $T_c$ and $10T_c$ were
numerically studied. Results were given for the $\omega$ dependence
of the spectral function at $k=0$, from which one could compute the $T$
dependence of the imaginary time correlator $G(\tau;T)$, measured on the lattice.
The real static correlator at $\omega=0$ was also computed and related to
structure observed near the light cone, $\omega\approx k$. Overall,
at $10T_c$ the correlators were very close to the conformal situation \cite{kv}, but
large effects were observed when $T$ decreased towards $T_c$.

The gravity background used here is IHQCD as developed 
in \cite{kiri1}-\cite{kiri4} for $k=0$
and a simplified and quasianalytically tractable
version thereof for all $\omega,k$. IHQCD describes very successfully
the properties of large $N_c$ Yang-Mills theory both at $T=0$ and at finite $T$,
but its gravity background, in general, has to be evaluated numerically. The
model used here retains the attractive features of IHQCD in the infrared, so that
it has confinement, glueballs and a first order phase transition, while simplifying
the complicated logarithmic structure of IHQCD in the ultraviolet. The chief
virtue of the model is that both $b(z)$ and $f(z)$, appearing in the fluctuation
equation, are analytically known. Another analytic approximation to IHQCD
\cite{aks,zahed}, would be quantitatively more accurate but does not permit
analytic evaluation of $f(z)$. In any case, we expect that the results of this
paper should be qualitatively valid for IHQCD, too, for all $\omega,k$.
This is confirmed by numerical computations at $k=0$.

There is ample phenomenological evidence for the strongly interacting nature of
QCD matter at $T\gsim T_c$ ($p={\fra34}p_\rmi{ideal}$, small viscosity, jet quenching),
but one has not been able to develop a first-principle numerically verifiable criterion
for this. We have suggested that a quantitative verification of our predictions could
serve this purpose.

There are several obvious directions of further study. For the first, a full numerical
computation in IHQCD should be carried out, also for nonzero $k$.
This is technically complicated due to
the fact that the expansions near the boundary are not power series in $z$ but contain
powers of $\log z$, studied here in Appendix A. Secondly, to extend the calculation to
correlators of the scalar operator $\tr F_{\mu\nu}^2$ one should use the more complicated
fluctuation equation in \cite{springer2}, coupling the dilaton and the
scalar component of the tensor fluctuation. A third open problem
is the computation of more terms in the large $\omega$ expansion \nr{imgasfull}.
Finally, it will be interesting to compare the results with those from
next-to-leading order perturbative QCD, when they are available.

\vspace{1cm}
{\it Acknowledgements}.
We thank J. Alanen, Jorge Casalderrey-Solana, Mikko Laine
and Fran\-cesco Nitti for discussions and advice.
F. Nitti has given us the Mathematica code used in \cite{kiri4}.
The work of MV has been supported by Academy of Finland, contract no. 128792 and
that of MK and AV by the Sofja Kovalevskaja program of the
Alexander von Humboldt foundation. KK and AV thank
the 2010 ESI workshop "AdS Holography and Quark-Gluon Plasma" in Vienna for hospitality.

\vspace{2cm}
\appendix
\section{Appendix: UV expansion in IHQCD}
In this appendix, we shall give small-$z$ logarithmic expansions
of $b(z)$, $\lambda(z)$ and $W(z)\equiv -\dot b/b^2$,
adding a few terms to those given in Appendix A of \cite{kiri1}.
Also $f(z)$ will obtain similar logarithmic corrections to the $z^4$ term
by integration of \nr{fdoteq}.

The expansions are given as powers of $\lambda(z)$ or as inverse powers of
\be
L\equiv\log{1\over\Lambda z},
\ee
where $\Lambda$ is the integration constant of
\be
{d\lambda\over dz}=-\beta W b(z).
\la{lambdaz}
\ee
The correction terms arise from the UV beta function expansion written in the form
\ba
b_0\beta(\lambda)&=&-(b_0\lambda)^2-{b_1\over b_0^2}(b_0\lambda)^3-{b_2\over b_0^3}(b_0\lambda)^4-
{b_3\over b_0^4}(b_0\lambda)^5+..,
\la{betexp}
\ea
where only $b_0,\,b_1$ are scheme independent.
It will appear that all quantities are functions of
\be
b_0\lambda,\qquad {b_i\over b_0^{i+1}}, \quad b\equiv {b_1\over b_0^2},
\ee
($b\equiv b_1/b_0^2$ is a standard notation, not to be confused with $b(z)$).

First, by integrating $db/b=d\lambda/\beta(\lambda)=db_0\lambda/(b_0\beta(\lambda))$,
(note the different constants $\hat b_0$ and $b_0$)
\be
{b\over\hat b_0}=\exp\left({1\over b_0\lambda}\right)\,\,(b_0\lambda)^{b_1/b_0^2}\left\{1+
\left({b_2\over b_0^3}-b^2\right)b_0\lambda+
\left[b(1+b)(b^2-2{b_2\over b_0^3})+{b_2^2\over b_0^6}+{b_3\over b_0^4}\right]{1\over2}
(b_0\lambda)^2+..\right\}.
\ee
The constant of integration $\hat b_0$ is fixed so that $b_0\lambda$ appears in
the powerlike term.

For $W\equiv -\dot b/b^2$ one can derive
\be
W(\lambda)=W(0)\exp\left(-\fra49\int_0^\lambda d\bar\lambda{\beta(\bar\lambda)
\over\bar\lambda^2}\right),\quad W(0)={1\over\CL}.
\la{W}
\ee
Expanding this one has
\be
\CL W=1+\fr49b_0\lambda+\left(\fr8{81}+\fr29{b_1\over b_0^2}\right)(b_0\lambda)^2+
4\left({8\over 3^7}+{2\over81}{b_1\over b_0^2}+
\fr1{27}{b_2\over b_0^3}\right)(b_0\lambda)^3+..
\la{Wexp}
\ee

The $z$ dependence is obtained by combining the equations $W=-\dot b/b^2$ and
$\beta=b\dot\lambda/\dot b$ into \nr{lambdaz}
so that, integrating,
\be
z=\int_0^z dz=\int_0^{\lambda(z)}{d\lambda\over -\beta W b}.
\la{zeq}
\ee
Noting that $W$ and $b$ contain
the constants $1/\CL$ and $\hat b_0$ this equation naturally defines the
scale
\be
\Lambda={\hat b_0\over\CL}
\ee
of $z$. Eq.~\nr{zeq} now has to be inverted for $\lambda$: one
expands the RHS around $\lambda=0$ using the expansions above, integrates term by
term and takes the limit $\lambda\to0$. Finally, the result is inverted recursively to
obtain $\lambda=\lambda(z)$. The expansion one obtains is
\ba
b_0\lambda&=&{1\over L}-{b\log L\over L^2}+
\left(b^2 \log^2L-b^2\log L+\fr49-b^2+{b_2\over b_0^3}\right){1\over L^3}+\nn
&&\hspace{-2cm}+\left[-b^3\log^3L+\fr52b^3\log^2L-
\left(3b{b_2\over b_0^3}-2b^3+\fr43 b\right)\log L
-\fr49+\fr23 b-\fr12 b^3+{b_3\over 2b_0^4}\right]{1\over L^4}+\nn
&&\hspace{-2cm}+\left[b^4\log^4 L-\fr{13}3 b^4\log^3 L+
\left(\fr32b^4+\fr83b^2+6b^2{b_2\over b_0^3}\right)\log^2 L+
\left(4b^4-4b^2+\fr{16}9b+3b^2{b_2\over b_0^3}-2b{b_3\over b_0^4}\right)\log L \right.\nn
&&\hspace{-2cm}\left. +\fr{104}{81}-\fr43b-\fr{10}9b^2+\fr76b^4+\fr{52}{27}{b_2\over b_0^3}
-3b^2{b_2\over b_0^3}+\fr53{b_2^2\over b_0^6}-\fr16b{b_3\over b_0^4}+\fr13{b_4\over b_0^5}
\right]{1\over L^5}+\CO({1\over L^6}).
\la{lamL}
\ea

To get corrections to $b(z)$ one writes \nr{lambdaz} in the form
\be
b(z)={\lambda'(z)\over-\beta W},\qquad \lambda'(z)=-{1\over z}{d\lambda\over dL},
\ee
uses \nr{lamL} and expands in $\lambda$.
This leads to 
\ba
b(z)&=&{\CL\over z} \left[1-{4\over9}b_0\lambda+
\left({44\over81}-{2\over9}b\right)(b_0\lambda)^2-
\left({2408\over2187}-{80\over81}b+{4\over27}{b_2\over b_0^3}\right)(b_0\lambda)^3+
\CO(\lambda^4)\right]\nn
&=&
{\CL\over z}\left\{1-{4\over9L}+{2(18b\log L+22-9b)\over81L^2}\,\,+\right.\la{bexp}\\
&&\hspace{-1cm}\left.
-4\left[{710\over2187}-{20\over81}b-\fr19b^2+\fr4{27}{b_2\over b_0^3}+
\left({22\over81}b-{2\over9}b^2\right)\log L+\fr19 b^2\log^2 L\right]{1\over L^3}+
\CO({1\over L^4})\right\}.\nonumber
\ea
The three first terms in the second form were given in (A.28) of \cite{kiri1}.

Finally, the UV expansion of the potential V is very simply directly obtained by
inserting the expansions of $W$ and $\beta$ to
$V(\lambda)=12W^2(\lambda)[1-\left(\beta/(3\lambda)\right)^2]$:
\be
V={12\over\CL^2}\left[1+\fr89 b_0\lambda+\left({23\over81}+\fr49b\right)(b_0\lambda)^2+
\left({40\over 2187}+{14\over81}b+{8\over27}{b_2\over b_0^3}\right)(b_0\lambda)^3+\CO(\lambda^4)
\right].
\la{Vexp}
\ee

\section{Appendix: IR expansion}
One cannot uniquely solve the equations \nr{eqf} at large $z$, large $\lambda$, but assume
that $b(z)$ behaves as
\be
b(z)=b_0 e^{-(\Xi z)^\alpha}(\Xi z)^p,\qquad \alpha\ge1,\quad p\,\,{\rm real}.
\ee
$\Xi$ here in the IR is a number times $\Lambda$ in the UV and, to emphasize
this difference, we use a different letter for it.
Inserting this to the equations of motion above and evaluating successively
$\dot b/b, bW, W, \dot W/W$ one can write $\dot\phi$ in a form which is
easily integrable. If $\lambda_0$ is the constant of integration of this equation,
one finds that, up to corrections $1+\CO(1/z^\alpha)\sim1+\CO(1/\log\lambda)$,
\ba
b(z)&=&b_0 e^{-(\Xi z)^\alpha} (\Xi z)^p=
{b_0\over\lambda^{2/3}}(\fra23\log\lambda)^{(\alpha-1)/(2\alpha)},\\
W(z)&=&\fra{\alpha}{\CL}e^{(\Xi z)^\alpha}(\Xi z)^{\alpha-1-p}
=\fra{1}{\CL}\lambda^{2/3}\alpha(\fra23\log\lambda)^{(\alpha-1)/(2\alpha)},
\quad \CL=\fra{b_0}{\Lambda},\\
\lambda(z)&=&e^{3(\Xi z)^\alpha/2}(\Xi z)^{3(\alpha-1-2p)/4},\nn
&&
\, e^{(\Xi z)^\alpha}=\lambda^{2/3}(\Lambda z)^{p-\alpha/2+1/2},\,\,
\Xi z=(\fra23\log\lambda)^{1/\alpha},\\
\beta(\lambda)&=&-\fra32\lambda
\left(1+{\alpha-1\over2\alpha(\Xi z)^\alpha}+\CO({1\over z^{2\alpha}})\right)=
-\fra32\lambda\left(1+{3(\alpha-1)\over4\alpha\log\lambda}+\CO({1\over\log^2\lambda})\right),\\
V(z)&=&9W^2(z)=\fra{9\alpha^2}{\CL^2}e^{2(\Xi z)^\alpha}(\Xi z)^{2(\alpha-1-p)}
=\fra{9\alpha^2}{\CL^2}\lambda^{4/3}(\fra23\log\lambda)^{(\alpha-1)/(\alpha)},
\la{largez}
\ea
where always $\lambda\equiv\lambda/\lambda_0$, apart from the leading term in $\beta(\lambda)$,
where the leading term is $-3\lambda/2$ for any $\lambda_0$.
By choosing $p$ suitably one can make any of the
quantities a pure exponential in $z^\alpha$.
However, when expressed in terms of $\lambda$, the leading
behavior is always independent of $p$. Note that $b\lambda^{2/3}\equiv b_s\sim
(\log\lambda)^{(\alpha-1)/(2\alpha)}$ is the string frame $b$,
which is used to express the confinement criterion.
Note also that the confining beta
function is not assumed here, it follows from the assumed form of $b(z)$.
The scalar potential $V$ satisfies $V=12W^2(1-X^2)$, $X\equiv\beta/(3\lambda)$.

Finally, Eq. \nr{fdoteq} now is
\be
\dot f(z)=C\,e^{3(\Xi z)^\alpha}(\Xi z)^{-3p}.
\ee
In our model with $\alpha=2,\,p=-1$ this is applies at all $z$
and is simply integrable in closed form.
If one wants just a large $z$ approximation for $f(z)$ one can integrate it up to
corrections of order $1/z^\alpha$ by extending the
integration to $z=0$ and imposing the condition $f(0)=1$. This leads to
\ba
f(z)&=&1-\left({z\over z_h}\right)^{1-\alpha-3p}e^{3\Xi^\alpha(z^\alpha-z_h^\alpha)},\nn
-\dot f(z_h)&=&4\pi T=3\alpha\Xi\,(\Xi z_h)^{\alpha-1}.
\ea

\section{Ultraviolet limit of the static correlator}
\label{app:olver}

In this appendix we derive the high-$k$ limit the static correlator
$G(\omega=0,\mathbf{k};T)$. Eliminating the first derivative in \nr{PQ} by writing
\be
\psi(y)=\sqrt{b^3f\sqrt{y}}\,\cdot\phi(y)
\ee
and setting $\omega=0$, we have the equation
\begin{equation}
 \psi''(y) = \biggl[\frac{\hat k^2}{y\, f(y)}
    -\biggl({f'(y)\over2f(y)}\biggr)^2+\fr14+{1\over2y}+{3\over4y^2}\biggr]\psi(y)
 \equiv \bigl[ \hat k^2 F(y)+G(y) \bigr]\psi(y) = 0
\label{eq:static_de}
\end{equation}
We wish to solve this around $y=0$ for $\hat k \to \infty$. The systematic approach
given in \cite{olver} starts by defining
\begin{equation}
 \xi \equiv \int^y_0 \frac{dx}{\sqrt{x f(x)}}, \qquad
 \psi(y) = \left( \frac{d\xi^2}{dy}\right)^{-1/2}W(\xi).
\end{equation}
The idea with replacing $y$ by the new variable $\xi$ is that the resulting
equation is in a Schr\"odinger-like form in which only $k^2$ multiplies the
function to be solved. Introducing $W$ eliminates the first derivative
generated. The solution then is given in the form\footnote{For the
coefficients $A_s$ and $B_s$, we follow here Exercise 5.2 of Chapter 12 in \cite{olver}.}
\begin{equation}
 W(\xi) = \xi K_2(\hat k\xi) \sum_{s=0}^\infty \frac{A_s(\xi)}{\hat k^{2s}}
  -\frac{\xi^2}{\hat k} K_3(\hat k\xi) \sum_{s=0}^\infty \frac{B_s(\xi)}{\hat k^{2s}}\, ,
\end{equation}
where the coefficients $A_s$ and $B_s$ are computed from the corresponding expansion
coefficients $\mathsf{A_s}$ describing the solution away from the singular point $\xi=0$:
\begin{align}
 A_s(\xi) &= \sum_{j=0}^{2s} a_j(3)\xi^{-j} \mathsf{A}_{2s-j}(\xi), &
 B_s(\xi) &= \sum_{j=0}^{2s+1} a_j(2)\xi^{-j-1} \mathsf{A}_{2s-j-1}(\xi),
\end{align}
where $a_j(\nu) = \prod_{k=1}^j (4\nu^2-(2k-1)^2)/8k$. Functions $\mathsf{A}_s$
in turn are given by
\begin{align}
 \mathsf{A}_{s+1}(\xi) &= -\frac{1}{2} \mathsf{A}_s'(\xi)+\frac{1}{2} \int \!d\xi\,
 \Psi(\xi)\mathsf{A}_s(\xi) \nonumber \\
    &= -\frac{1}{2} \sqrt{y\,f(y)}\mathsf{A}_s'(y)
	+\frac{1}{2} \int \! \Psi(y)\mathsf{A}_s(y) \frac{dy}{\sqrt{y\,f(y)}}.
\end{align}
The integration constants in each step are chosen so that when we retrace our
steps back to the original mode function $\phi(y)$, we have $\phi(0)$ finite
and equal to 1. The function $\Psi(y)$ in the above recursion relation is
given by the coefficient functions $F(y),G(y)$ of eq.~(\ref{eq:static_de}), and reads
\begin{equation}
 \Psi(y) = \frac{G(y)}{F(y)}-F(y)^{-3/4} \frac{d^2}{dy^2}F(y)^{-1/4} \approx
    \frac{15}{16y}+\frac{1}{2} +\frac{ 8 y_h -8 +5e^{-y_h}}{32 (y_h-1+e^{-y_h})}y+\ldots
\end{equation}

Working out the series expansions of $\mathsf{A}_s$, $s=0\ldots 4$ (with
$\mathsf{A}_s$=constant), we can read out the coefficients of the $y^0$ and $y^2$ terms in
\begin{equation}
 \phi(y) = (b^3f\sqrt{y})^{-1/2}\psi(y) = (2\xi b^3\sqrt{f})^{-1/2}w(\xi)
\end{equation}
to finally arrive at
\begin{equation}
\frac{B}{A} = -\hat k^4\left(\ln \hat k +\gamma_E -\frac{3}{4}\right)
    -\hat k^2\left(\ln \hat k +\gamma_E \right)
    -\frac{5}{24} +{e^{-y_h}\over20(y_h-1+e^{-y_h})}\, .
    \la{careful}
\end{equation}

\end{document}